\documentclass[prb,amsmath,amssymb,showpacs,showkeys,twocolumn]{revtex4}
\usepackage{graphicx}
\usepackage{dcolumn}
\usepackage{rotating}
\usepackage{amssymb}
\usepackage{mathptmx}

\usepackage{amsfonts}
\usepackage{amsmath}
\usepackage{bm}
\begin{document}

\title {On the nonlinear NMR and magnon BEC in antiferromagnetic materials with coupled electron-nuclear spin precession}

\author{L.~V. Abdurakhimov$^{1}$}
\altaffiliation[Current address: ]{London Center for Nanotechnology, University College London, London WC1H 0AH, United Kingdom}
\author{M.~A. Borich$^{2,5}$}
\author{Yu.~M. Bunkov$^{1,3,4}$}
\author{R.~R. Gazizulin$^{3,4}$}
\author{D. Konstantinov$^{1}$}
\email{denis@oist.jp}
\author{M.~I. Kurkin$^{2}$}
\author{A.~P. Tankeyev$^{2,5}$}

\affiliation{$^{1}$ Okinawa Institute of Science and Technology (OIST) Graduate University, Onna, 904-0495 Okinawa, Japan\\
$^{2}$ Institute of Metal Physics, Ural Branch of Russian Academy of Sciences, 620990 Ekaterinburg, Russia\\
$^{3}$ CNRS Institute NEEL et Universite Grenoble Alpes, 38042 Grenoble, France\\
$^{4}$ Kazan  Federal University, 420008 Kazan, Russia\\
$^{5}$ Ural Federal University, 620002 Ekaterinburg, Russia}

\begin{abstract}

We present a new study of nonlinear NMR and Bose-Einstein Condensation (BEC) of nuclear spin waves in antiferromagnetic $\mathrm{MnCO_{3}}$ with coupled electron and nuclear spins. In particular, we show that the observed behaviour of NMR signals strongly contradicts the conventional description of paramagnetic ensembles of noninteracting spins based on the phenomenological Bloch equations. We present a new theoretical description of the coupled electron-nuclear spin precession, which takes into account an indirect relaxation of nuclear spins via the electron subsystem. We show that the magnitude of the nuclear magnetization is conserved for arbitrary large excitation powers, which is drastically different from the conventional heating scenario derived from the Bloch equations. This provides strong evidence that the coherent precession of macroscopic nuclear magnetization observed experimentally can be identified with BEC of nuclear spin waves with $\textbf{k}=0$.      

\end{abstract}

\date{\today}



\maketitle

\section{Introduction}

Bose-Einstein Condensation (BEC) of bosons, or accumulation of a macroscopic number of bosons in the same quantum state, presents one of the cornerstone subjects of studies in modern condensed-matter physics. In addition to the well-known phenomena of superfluidity, superconductivity, and BEC of cold atoms, the BEC of spin waves (magnons) in magnetically ordered materials has attracted a lot of recent attention.~\cite{book,Safonov,Vasiliev2015} Unlike conventional atomic BEC obtained by cooling atomic systems, magnon BEC can be established, for example, by continuous external rf pumping, which compensates for the loss of quasiparticles. Uniform precession of ordered spins in such systems can be described by magnons in a single quantum state with the wave number $k=0$, which thus can be called BEC of magnons in the  $\textbf{k}=0$ state. On the other hand, a qualitatively similar state of uniform precession exists for a paramagnetic system of noninteracting spins under resonant excitation by an external rotating magnetic field. Since such a system of non-interacting spins cannot be described in terms of magnons, it is important to establish a clear physical picture of the BEC of magnons with $k=0$ in magnetically ordered systems, and to distinguish it from uniform precession of noninteracting spins induced by external rf excitation.  
 
\begin{figure}[htt]
\includegraphics[width=0.44\textwidth]{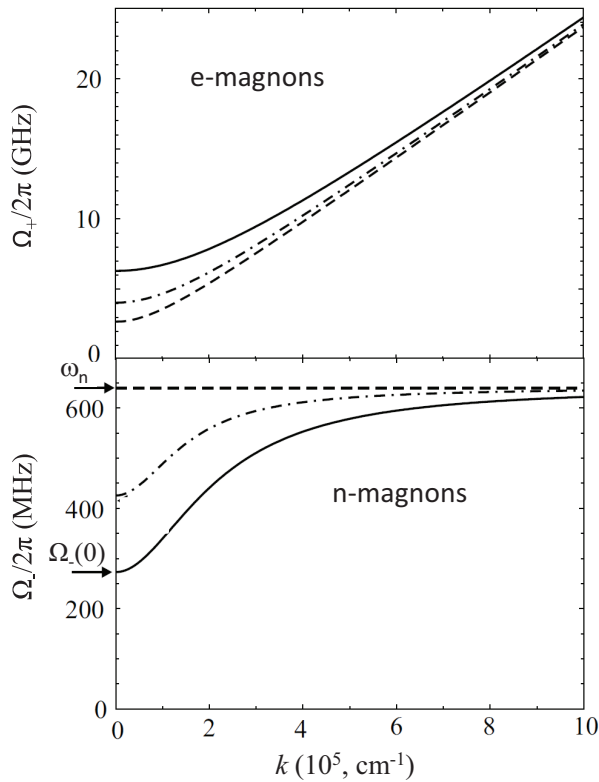}
\caption{ The spectra of e-magnons and n-magnons in $\mathrm{MnCO_{3}}$ in the static magnetic field of $0.2\,$kOe and at two values of temperature of $5\,$K (dashed-point line) and $1.4\,$K (solid line). The dashed lines show corresponding spectra in the case when coupling between electron and nuclear oscillations is neglected.}
\label{fig:1}
\end{figure}

In this work, we focus on magnetically ordered spin systems with coupled electron-nuclear spin precession. In such systems, dynamics of paramagnetic nuclear spins are governed by the exceptionally strong (10$^5$-10$^6\,$Oe) hyperfine field from ordered electron spins. In turn, motion of electron spins is affected by the hyperfine field from nuclear spins. This results in coupled electron-nuclear spin oscillations that can be described in terms of electron and nuclear spin waves.~\cite{deGennes,SN} In particular, hyperfine interaction leads to hybridization of, on the one hand, the usual spin waves in the magnetically ordered electron subsystem, i.e. e-magnons, and on the other hand, precession of a nuclear spin around the direction of the hyperfine field. Fig.~\ref{fig:1} shows examples of spectra for hybridized e-magnons (frequency $\Omega_{+}$) and n-magnons (frequency $\Omega_{-}$) in a weakly anisotropic antiferromagnet, $\mathrm{MnCO_{3}}$. At zero wave number, $k=0$, the frequency of n-magnons to a good approximation can be written as~\cite{TurovKuleev} 

\begin{equation}
\Omega_{-}(0)\approx \omega_n - \omega_p(H,T) \cos \beta,
\label{eq:bot}
\end{equation}

\noindent where $\omega_{n}$ is the frequency of precession of the nuclear spin in a stationary hyperfine field (that is when the oscillations of the electron spin system caused by coupling to the nuclear spins are ignored), $\omega_p$ is a quantity that depends on a number of parameters including the applied magnetic field $H$ and temperature of the nuclear system $T$, and $\beta$ is the angle of deflection of the nuclear magnetization vector from its equilibrium orientation $\textbf{z}$. The second term on the right-hand side of (\ref{eq:bot}) is usually known as "frequency pulling". It is clear that the frequency $\Omega_-(0)$ corresponds to the usual NMR frequency, which can be measured in the experiment by applying a uniform ac magnetic field $\textbf{h}(t)\perp \textbf{z}$. Note that at large $k$ the spectrum of the nuclear branch approaches $\omega_n$. Therefore, for the existence of $n$-magnons, the frequency difference between $\Omega_{-}(0)$ and $\omega_{n}$ must satisfy the condition

\begin{equation}
\omega_{n}-\Omega_{-}(0)\gg \delta\omega_{n},
\end{equation}

\noindent where $\delta\omega_{n}$ is the spread in the frequency $\omega_{n}$ due to inhomogeneities of the atomic and magnetic structures of the sample (inhomogeneous broadening) and due to relaxation processes (homogeneous broadening). There are few crystalline materials that satisfy this condition. Most notable examples are the weakly anisotropic antiferromagnets of the easy-plane type with $^{55}$Mn as the magnetic ions, such as $\mathrm{MnCO_{3}}$, $\mathrm{CsMnF_{3}}$, $\mathrm{RbMnF_{3}}$, $\mathrm{KMnF_{3}}$.  

The existence of the n-magnon branch with minima at $k=0$, see Fig.~\ref{fig:1}, suggests the possibility for BEC of nuclear magnons with $k=0$ under external rf pumping. Recent investigations in easy-plane, antiferromagnetic $\mathrm{CsMnF_{3}}$ and $\mathrm{MnCO_{3}}$ crystals using CW and pulsed NMR methods have shown evidence of magnon BEC.~\cite{Kazan1,Kazan2,Kazan3} In particular, the relatively strong and long-lived nuclear precession signals observed in pulsed NMR experiments indicated that nuclear magnetization is deflected by a large angle from the equilibrium orientation, while preserving its magnitude. This would strongly contradict the conventional behavior of a paramagnetic ensemble of noninteracting nuclear spins under external rf pumping, which is traditionally described by the phenomenological Bloch equations and predicts a strong reduction of the magnitude of the nuclear magnetization vector due to heating of the nuclear spin system. Thus, a detailed study and new, more  adequate description of coupled electron-nuclear spin systems under external rf pumping is necessary to establish a clear physical picture of magnon BEC in such systems.         

In this work, we present a detailed study of coupled electron-nuclear spin dynamics in easy-plane, weakly anisotropic, antiferromagnet $\mathrm{MnCO_{3}}$, under continuous NMR excitation. Our theoretical analysis shows that the coupling between electron and nuclear spin precession provides a mechanism not only for correlations between nuclear spins, that is, formation of n-magnons, but also relaxation of nuclear spins via the electron subsystem. This new mechanism of indirect relaxation preserves the magnitude of the nuclear magnetization vector for arbitrary large excitation powers; thus, it excludes the possibility of heating the nuclear subsystem by external pumping. This theory accounts well for the behavior of nonlinear NMR signals obtained in an antiferromagnetic $\mathrm{MnCO_{3}}$ sample at temperatures below $1\,$K. It is also shown that the conventional Bloch approach fails to account for our experimental results. Finally, we show that in the absence of heating, the energy of external pumping can be entirely transfered into the energy of uniform precession of nuclear spins, and the latter can be associated with BEC of n-magnons with $\textbf{k}=0$.

In Section II, we provide a macroscopic description of coupled electron-nuclear spin systems based on equations of motion linearized with respect to transverse components of the electron magnetization vector. In Section III, we describe measurements of nonlinear NMR signals in an antiferromagnetic $\mathrm{MnCO_{3}}$ sample and provide a detailed comparison with our theory. In Section IV, we discuss a microscopic picture of coupled electron-nuclear spin systems under excitation in terms of n-magnons and formation of a uniform precession of nuclear spins, which can be identified as magnon BEC. Some details for derivation of the coupled equations of motion for an electron-nuclear spin system are given in the Appendix.   
   
\section{Theoretical description}

Our theoretical analysis of a coupled electron-nuclear spin system in easy-plane, antiferromagnet MnCO$_3$ is based on the classical Landau-Lifshitz-Gilbert equations for macroscopic magnetization vectors of magnetic sub-lattices\cite{landau1,landau2,landau3,theory,tyablikov}. As usual, such an analysis of quantum spin systems is justified because the quantum-mechanical equations of motion reduce to classical equations for the precession of macroscopic magnetization vectors after averaging over the quantum-mechanical density matrix operator. Following the previous work~\cite{Turov}, we describe the dynamics of a coupled electron-nuclear spin system using equations for two electron ($\mathbf{M}_{1}$ and $\mathbf{M}_{2}$) and two nuclear ($\mathbf{m}_{1}$ and $\mathbf{m}_{2}$) magnetization vectors of two sub-lattices

\begin{eqnarray}
&& \frac{\mathrm{d} \mathbf{M}_{1}}{\mathrm{d} t}=\gamma _{e}\left ( \mathbf{M}_{1}\times \mathbf{H}_{\rm{M}_1} \right )+\mathbf{R}_{\rm{M}_1}, \nonumber \\ 
&& \frac{\mathrm{d} \mathbf{M}_{2}}{\mathrm{d} t}=\gamma _{e}\left ( \mathbf{M}_{2}\times \mathbf{H}_{\rm{M}_2} \right )+\mathbf{R}_{\rm{M}_2},
\label{eq:M}
\end{eqnarray}

\begin{eqnarray}
&& \frac{\mathrm{d} \mathbf{m}_{1}}{\mathrm{d} t}=\gamma _{n}\left ( \mathbf{m}_{1}\times \mathbf{H}_{\rm{m}_1} \right )+\mathbf{R}_{\rm{m}_1}, \nonumber \\
&& \frac{\mathrm{d} \mathbf{m}_{2}}{\mathrm{d} t}=\gamma _{n}\left ( \mathbf{m}_{2}\times \mathbf{H}_{\rm{m}_2} \right )+\mathbf{R}_{\rm{m}_2},
\label{eq:m}
\end{eqnarray}

\noindent where $\mathbf{H}_{\rm{M}_1}$ and $\mathbf{H}_{\rm{M}_2}$ ($\mathbf{H}_{\rm{m}_1}$ and $\mathbf{H}_{\rm{m}_2}$) are the effective magnetic fields acting on each electron (nuclear) sub-lattice, and  $\mathbf{R}_{\rm{M}_1}$ and $\mathbf{R}_{\rm{M}_2}$ ($\mathbf{R}_{\rm{m}_1}$ and $\mathbf{R}_{\rm{m}_2}$) are the relaxation terms for the electron (nuclear) magnetization in each sub-lattice. The effective magnetic fields are given by functional derivatives

\begin{equation}
\mathbf{H}_{\rm{M}_i}=-\frac{\delta \Phi }{\delta \mathbf{M}_i}, \quad \mathbf{H}_{\rm{m}_i}=-\frac{\delta \Phi }{\delta \mathbf{m}_i}, \quad i=1,2.
\end{equation}

\noindent The thermodynamic potential of the whole spin system, including its interaction with external static and rf magnetic fields is given by 

\begin{equation}
\Phi = \Phi_{ex} + \Phi_{A} + \Phi_{D} + \Phi_{H} + \Phi_{hyp}.\
\label{eq:Phi}
\end{equation}

\noindent The exchange interaction between electron spins of the two sub-lattices is described by

\begin{equation}
\Phi _{ex}=\frac{1}{V}\int d\boldsymbol{\mathbf{r}}\int d\mathbf{r'}J\left ( r-{r}' \right )\mathbf{M}_{1}\left ( r \right )\mathbf{M}_{2}\left ( {r}' \right ),
\end{equation}

\noindent where $J>0$ describes the exchange interaction and $V$ is the sample volume. The energy of magnetic anisotropy with a dedicated axis $\mathbf{c}$ chosen as the positive $\mathbf{y}$ direction is described by

\begin{equation}
\Phi _{A}=\frac{1}{V}\int d\mathbf{r}K\left [ \left ( M_{1}^{y}\left ( r \right ) \right )^{2}+\left ( M_{2}^{y}\left ( r \right ) \right )^{2} \right ].
\end{equation}

\begin{figure}[htt]
\includegraphics[width=0.50\textwidth]{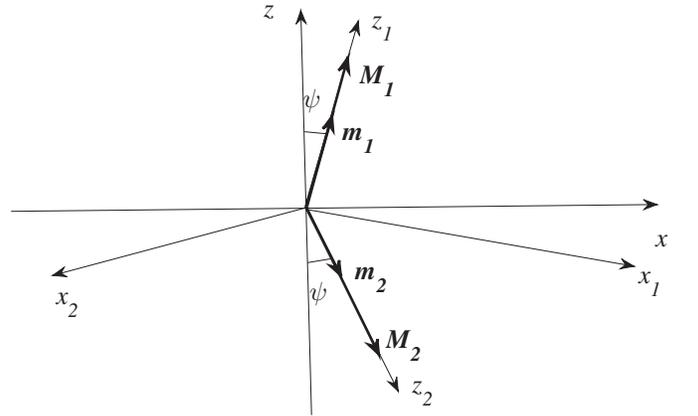}
\caption{Equilibrium position of the electron and nuclear magnetization vectors for each sub-lattice.}
\label{Fig:Geom}
\end{figure}

\noindent In the antiferromagnet MnCO$_{3}$ considered here, this anisotropy is the easy-plane that corresponds to the condition $K>0$. The Dzyaloshinskii-Moriya interaction which is responsible for non-collinearity of the $\mathbf{M}_{1}$ and $\mathbf{M}_{2}$ sub-lattices, see Fig.~\ref{Fig:Geom}, is described by

\begin{equation}
\Phi _{D}=\frac{1}{V}\int d\mathbf{r}D\left ( M_{1}^{x}\left ( r \right ) M_{2}^{z}\left ( r \right )-M_{1}^{z}\left ( r \right )M_{2}^{x}\left ( r \right )\right ).
\end{equation}

\noindent where $D$ is a constant that characterizes the strength of the Dzyaloshinkii-Moriya interaction. The energy of $\mathbf{M}_{1}$ and $\mathbf{M}_{2}$ magnetizations in the external static magnetic field $\mathbf{H}$ applied in the positive $\mathbf{x}$ direction and the ac magnetic field  $\mathbf{h}(t)$ applied along axis $z$ is described by

\begin{eqnarray}
&& \Phi _{H}=-\frac{1}{V}\int d\mathbf{r}H\left ( M_{1}^{x}\left ( r \right )+M_{2}^{x}\left ( r \right ) \right )- \nonumber \\
&& - \frac{1}{V}\int d\mathbf{r}h\left ( t \right )\left ( M_{1}^{z}\left ( r \right )+M_{2}^{z}\left ( r \right ) \right ).
\label{eq:PhiH}
\end{eqnarray}

\noindent In what follows, we neglect the direct interaction of nuclear magnetization vectors $\bf{m}_1$ and $\bf{m}_2$ with the static and rf magnetic fields compared to the much stronger hyperfine field. However, note that the rf magnetic field still strongly couples to the nuclear spins via the $\mathbf{h}(t)$-induced oscillations of electron magnetization. The hyperfine interaction between electron and nuclear spins, which for simplicity we assume to be isotropic, is described by
 
\begin{equation}
\Phi _{hyp}=-\frac{1}{V}\int d\mathbf{r}A\left ( \mathbf{M}_{1}\left ( r \right )\mathbf{m}_{1}\left ( r \right )+\mathbf{M}_{2}\left ( r \right )\mathbf{m}_{2}\left ( r \right ) \right ).
\label{eq:PhiHyp}
\end{equation}

At $\mathbf{h}(t)$=0, the minimum of the functional (\ref{eq:Phi}) defines the equilibrium orientation of the vectors $\mathbf{M}_{1,2}$ and $\mathbf{m}_{1,2}$, see Fig.~\ref{Fig:Geom}. The small angle $\psi$ in this figure is determined from the equation

\begin{equation}
\sin \psi = \frac{H+H_{D}}{H_{E}},
\label{eq:sin}
\end{equation}

\noindent where $H_{D}=DM_{0}$ is the magnitude of the Dzyaloshinskii-Moria field, $M_{0}=\left |  \boldsymbol{\mathbf{M}}_1\right |=\left |  \boldsymbol{\mathbf{M}}_2\right |$ is the magnitude of the electron magnetization vector in each sub-lattice, and the exchange field is given by

\begin{equation}
H_{E}=2M_{0}\int d\left ( \mathbf{r-r'} \right )J\left ( \mathbf{r-r'} \right )
\end{equation}

\noindent The typical angle $\psi$ is on the order of $10^{-2}$ rad.

The solutions of Eqs.~(\ref{eq:M}-\ref{eq:PhiHyp}) for small deviations of $\mathbf{M}_{1,2}$ and $\mathbf{m}_{1,2}$ from their equilibrium orientations were investigated earlier.~\cite{deGennes,Turov} In this case, the eigenmodes of coupled electron-nuclear oscillations can be found from Eqs.~(\ref{eq:M}-\ref{eq:m}) linearized with respect to the transverse components of magnetization vectors $\mathbf{M}_{1,2}$ and $\mathbf{m}_{1,2}$ and by taking the rf field $\textbf{h}(t)$ and relaxation terms to be zero. As expected, the corresponding eigenfrequencies coincide with frequencies of coupled electron-nuclear spin waves at $k=0$, see Eq.~(\ref{eq:bot}). Electron and nuclear ac susceptibilities can be obtained from the same linearized equations by including the homogeneous rf magnetic field $\textbf{h}(t)$.~\cite{TurovKuleev}    

In what follows, we are interested in finding solutions of Eqs.~(\ref{eq:M}-\ref{eq:PhiHyp}) for arbitrarily large deviations of vectors $\mathbf{m}_{1,2}$ from their equilibrium orientations due to the rf magnetic field at frequencies close to the NMR resonance. Following the previous work,~\cite{theory} it is convenient to introduce a separate reference frame $x_1y_1z_1$ ($x_2y_2z_2$) for sublattice $\bf{M}_1$ ($\bf{M}_2$) obtained from the original frame $xyz$ by a clockwise rotation around the $z$-axis by angle $\psi$ (by angle $\pi-\psi$), see Fig.~\ref{Fig:Geom}, and we chose new variables according to

\begin{eqnarray}
&& 2M^{\xi}=M_1^{x_1} + M_2^{x_2}, \quad 2m^{\xi}=m_1^{x_1} + m_2^{x_2}, \nonumber \\
&& 2M^{\eta}=M_1^{y_1} + M_2^{y_2}, \quad 2m^{\eta}=m_1^{y_1} + m_2^{y_2}, \nonumber \\
&& 2M^{\zeta}=M_1^{z_1} + M_2^{z_2}, \quad 2m^{\zeta}=m_1^{z_1} + m_2^{z_2},
\label{eq:magMm}
\end{eqnarray}  

\noindent where $M_i^{x_i}$, $M_i^{y_i}$, $M_i^{z_i}$ ($m_i^{x_i}$, $m_i^{y_i}$, $m_i^{z_i}$), $i=1,2$, are components of electron (nuclear) magnetization vectors in the corresponding frame. The solutions of Eqs.~(\ref{eq:M}-\ref{eq:m}) in these new variables can by found by linearizing Eqs.~(\ref{eq:M}) with respect to transverse components of electron magnetization vectors, whose deviation from the equilibrium orientation is assumed to be small. This assumption is valid for sufficiently large magnetic fields $H>H_c$~\eqref{A2} where the oscillation of electron spins is only weakly affected by the hyperfine interaction with nuclei, see discussion in the previous section. The details of its derivation are given in the Appendix. Finally, is is convenient to write the result in the coordinate system $XYZ,$ where the components of the net nuclear magnetization $m_X$, $m_Y$ and $m_Z$ are related to $m^{\xi}$, $m^{\eta}$ and $m^{\zeta}$ by

\begin{eqnarray}
&& m^{\xi}=m_X\cos \omega t + m_Y \sin \omega t, \nonumber \\
&& m^{\eta}=-m_X\sin \omega t + m_Y \cos \omega t, \nonumber \\
&& M^{\zeta}=m_Z,
\label{eq:RWT}
\end{eqnarray}
   
\noindent which is somewhat similar to the usual rotating wave transformation used to describe traditional NMR.~\cite{Abragam}  

By omitting relaxation terms in Eqs.~(\ref{eq:M}-\ref{eq:m}) equations for the net nuclear magnetization become (see Appendix for details) 

\begin{eqnarray}
&&\frac{\mathrm{d}m_{X}}{\mathrm{d}t}=\left(\omega_{n}-\omega-\omega_{p}\frac{m_{Z}}{m_{0}}\right)m_{Y},
\label{eq:mXnoR}\\
%
&&\frac{\mathrm{d}m_{Y}}{\mathrm{d}t}=-\left(\omega_{n}-\omega-\omega_{p}\frac{m_{Z}}{m_{0}}\right)m_{X}+\omega_{1}m_{Z},
\label{eq:mYnoR}\\
%
&&\frac{\mathrm{d}m_{Z}}{\mathrm{d}t}=-\omega_{1}m_{Y},
\label{eq:mZnoR}
\end{eqnarray}

\noindent where $\omega_1$ is proportional to the amplitude of the rf magnetic field $\bf{h}_1$ and $\omega_p$ is the parameter discussed previously (see Eqs.~\eqref{A10},\eqref{A12}, in the Appendix for the precise definition of $\omega_1$ and $\omega_p$). Since the relaxation terms were neglected, the above equations can be used to describe the dynamics of the magnetization vector $\bf{m}$ only on time intervals $t$, which are small compared to the nuclear magnetic relaxation time. Note that as was discussed previously from the microscopic point of view, the presence of the frequency pulling term $\omega_p(m_Z/m_0)$ does not allow us to obtain large deflection angles of $\bf{m}$ from the equilibrium orientation using a short rf pulse, as can typically be done in conventional NMR (the Rabi flopping).~\cite{Abragam} Indeed, for typical experimental conditions we have $\omega_p>>\omega_1$, therefore the orientation of the axis of rotation of vector $\bf{m}$ in the frame $XYZ$ changes quickly to be nearly parallel to $\bf{Z}$, after applying the rf pulse. However, as we show below, it is possible to obtain large angles  of deflection by CW NMR. The description of the stationary state of the spin system under CW rf pumping requires accounting for the proper form of relaxation terms in Eqs.~(\ref{eq:M}-\ref{eq:m}). 

As discussed previously, interaction between transverse components of the electron and nuclear magnetization vectors significantly changes the frequency of nuclear spin oscillations. Another consequence of this interaction is the appearance of the relaxation of the transverse component of vector $\bf{m}$. Indirect relaxation of this type has been studied previously, but for relatively small amplitudes of $\bf{m}$-vector oscillations when nonlinear phenomena were insignificant.~\cite{Turov} Here, we consider nonlinear equations of motion for nuclear magnetization taking into account its indirect relaxation via the electron spin system. 

Typically, relaxation of the transverse component of magnetization occurs due to fluctuating magnetic fields acting on different spins, and to inhomogeneity of the external static magnetic field acting in the electron system. In what follows, we use an assumption that, because nuclear spins experience an enormously large hyperfine field $H_e\sim 10^5$-$10^6\,$Oe, we can neglect any other fields, including fluctuating magnetic fields responsible for relaxation of vectors $\bf{M}_{1,2}$. We note that the approximation $\bf{R}_{\textrm{m}_{1,2}}$=0 in Eqs.~(\ref{eq:m}) conserves the magnitude of vectors $\bf{m}_{1,2}$ and the net magnetization vector $\bf{m}$. Thus, this approximation excludes the possibility of heating of the nuclear spin system by any processes, including its excitation by rf pumping. 

Using this approximation together with the assumption of small deviations of electron magnetization vectors from their equilibrium orientations at $H>H_c$, equations of motion for the vector $\bf{m}$ become (see Appendix for details)      

\begin{eqnarray}
&&\frac{\mathrm{d}m_{X}}{\mathrm{d}t}=\left( \omega_{n}-\omega-\omega_{p}\frac{m_{Z}}{m_{0}}\right)m_{Y}-\frac{m_{Z}}{m_{0}}\frac{m_{X}}{T_{2n}},
\label{eq:mXLLG}\\
%
&&\frac{\mathrm{d}m_{Y}}{\mathrm{d}t}=-\left(\omega_{n}-\omega-\omega_{p}\frac{m_{Z}}{m_{0}}\right)m_{X} + \omega_{1} m_{Z} - \frac{m_{Z}}{m_{0}}\frac{m_{Y}}{T_{2n}},\qquad
\label{eq:mYLLG}\\
%
&&\frac{\mathrm{d}m_{Z}}{\mathrm{d}t}=-\omega_{1}m_{Y}-\frac{m_{Z}+m_{0}}{m_{0}}\frac{m_{Z}-m_{0}}{T_{2n}},
\label{eq:mZLLG}
\end{eqnarray}

\noindent where the nuclear transverse relaxation time $T_{2n}$ is related to the electron transverse relaxation time $T_2e$ by Eq.~(\ref{A13}) in the Appendix.

CW NMR signals are described by stationary state solutions of the above equations. The latter are given by 

\begin{eqnarray}
&&\frac{m_{X}}{m_{0}}=\frac{\left ( \omega_{n}-\omega-\omega_{p}+\omega_{p}\Delta \right ) \omega_{1}\left ( 1-\Delta \right )}{\left (\omega_{n}-\omega-\omega_{p}+\omega_{p}\Delta \right )^{2}+\Gamma_{1}^{2}},
\label{eq:mXstead}\\
%
&&\frac{m_{Y}}{m_{0}}=\frac{\omega_{1}\Gamma_{1}\left ( 1-\Delta \right )}{\left( \omega_{n}-\omega-\omega_{p}+\omega_{p}\Delta\right)^{2}+\Gamma_{1}^{2}},
\label{eq:mYstead}
\end{eqnarray}

\noindent where $\Delta=1-m_{Z}/m_{0}$, which determines that the longitudinal component of $\bf{m}$, is the root of the fourth degree polynomial equation

\begin{equation}
\Delta(2-\Delta)=\frac{\omega_{1}^2\left( 1-\Delta \right)^2}{\left( \omega_{n}-\omega-\omega_{p}+\omega_{p}\Delta\right )^{2}+\Gamma_1^2},
\label{eq:mZstead}
\end{equation}

\noindent and $\Gamma_1$ is given by

\begin{equation}
\Gamma_1=\left( 1-\Delta \right)/T_{2n}.
\label{eq:Gamma1}
\end{equation}

For the sake of qualitative comparison with our theory, let us also consider the traditional Bloch approach in which the nuclear relaxation is described by introducing two phenomenological relaxation times, i.e. the longitudinal magnetization relaxation time $T_1$ and the transverse magnetization relaxation time $T_2$.~\cite{Abragam} Adding the corresponding relaxation terms into Eqs.~\eqref{eq:mXnoR}-\eqref{eq:mZnoR}, we obtain the phenomenological Bloch equations  

\begin{eqnarray}
&&\frac{\mathrm{d}m_{X}}{\mathrm{d}t}=\left( \omega_{n}-\omega-\omega_{p}\frac{m_{Z}}{m_{0}}\right)m_{Y}-\frac{m_{X}}{T_2},
\label{eq:mXBloch}\\
%
&&\frac{\mathrm{d}m_{Y}}{\mathrm{d}t}=-\left(\omega_{n}-\omega-\omega_{p}\frac{m_{Z}}{m_{0}}\right)m_{X} + \omega_{1} m_{Z} - \frac{m_{Y}}{T_2},
\label{eq:mYBloch}\\
%
&&\frac{\mathrm{d}m_{Z}}{\mathrm{d}t}=-\omega_{1}m_{Y}-\frac{(m_{Z}-m_{0})}{T_1}.
\label{eq:mZBloch}
\end{eqnarray}

\noindent Note that the above equations are obtained by disregarding the electron spin relaxation terms $\mathbf{R}_{\rm{M}_1,2}$ in Eqs.~(\ref{eq:M}). The stationary state solutions of (\ref{eq:mXBloch}-\ref{eq:mZBloch}) can be written as

\begin{eqnarray}
\frac{m_{X}}{m_{0}}=\frac{\left( \omega_{n}-\omega-\omega_{p}+\omega_{p}\Delta \right) \omega_{1} ( 1-\Delta )}{\left(\omega_{n}-\omega-\omega_{p}+\omega_{p}\Delta \right)^{2}+ T_2^{-2}},
\label{eq:mXsteadBloch}\\
%
\frac{m_{Y}}{m_{0}}=\frac{\omega_{1}T_2^{-1}( 1-\Delta )}{\left(\omega_{n}-\omega-\omega_{p}+\omega_{p}\Delta \right)^{2}+ T_2^{-2}},
\label{eq:mYsteadBloch}
\end{eqnarray}

\noindent where $\Delta=1-m_Z/m_0$, as defined earlier, is given by the roots of the third degree polynomial equation

\begin{equation}
\Delta=\frac{\omega_{1}^2T_1T_2^{-1}( 1-\Delta )}{\left(\omega_{n}-\omega-\omega_{p}+\omega_{p}\Delta \right)^{2} + T_2^{-2}}.
\label{eq:mZsteadBloch}
\end{equation}

To distinguish between the two approximations which lead, on the one hand, to Eqs.~(\ref{eq:mXstead}-\ref{eq:mZstead}), and on the other hand, to Eqs.~(\ref{eq:mXsteadBloch}-\ref{eq:mZsteadBloch}), we need to compare the continuous NMR signals predicted by these equations with experimentally observed signals. This is done in the next section.  

\section{Experimental Comparisons}
\subsection{Experimental methods}

In the experiment, we studied CW NMR signals in an easy-plane antiferromagnetic $\mathrm{MnCO_{3}}$ sample at temperatures below $1\,$K. The sample was in the form of a rhombus-shaped plate about $0.7\,$mm thick, with diagonals of approximately $2.4$ and $2.7\,$mm. The magnetic anisotropy axis $\bf{c}$ was perpendicular to the plane of the plate. The measured mass of the sample was about $8\times 10^{-3}\,$g.

To excite continuous NMR signals in $\mathrm{MnCO_{3}}$ by external rf pumping, the sample was placed in a high-quality factor rf resonator of a split-ring type.~\cite{HardySR} The resonator and sample were oriented such that the external static magnetic field $\bf{H}$ and the rf magnetic field $\bf{h}$ of the resonator were perpendicular to each other and to the anisotropy axis $\bf{c}$ of the sample. The resonator was made of high-conductivity, oxygen-free copper and had a resonant frequency of about $591.3\,$MHz and a quality factor of about 600 measured at $1\,$K. The input rf signal transmitted through a cryogenic, low-loss, semi-rigid coaxial cable was coupled to the resonator via a loop made at the end of the cable, and the reflected signal was measured at different values of static field $H$ and input rf power $P$ using a network analyzer. Signals proportional to the in-phase and quadrature (with respect to the rf field in the cavity) components of ac magnetization induced in the sample could be extracted from the reflected signal. In order to avoid complications associated with the strong coupling of the spin system to the electro-magnetic mode of the resonator, which lead to normal mode splitting in the resonator reflection spectrum,~\cite{Abdur2015} we used the frequency of rf pumping of $593.5\,$MHz, which is significantly detuned from the resonant frequency of the resonator. In this so called dispersive regime of coupling, the in phase ($M_d$) and quadrature ($M_a$) components of the ac magnetization are related to the complex amplitude (phasor) of the reflected signal, according to

\begin{equation}
V\approx V_0 + C(-M_a + iM_d),
\label{eq:exp}
\end{equation}  

\noindent where $V_0$ is the complex amplitude of the signal reflected from an empty cavity, $i$ is imaginary unit, and $C>0$ is a proportionality coefficient that depends on characteristics of the resonator and its coupling to the rf transmission line.        
%
%

\subsection{Experimental results}

\begin{figure}[htt]
\includegraphics[width=0.48\textwidth]{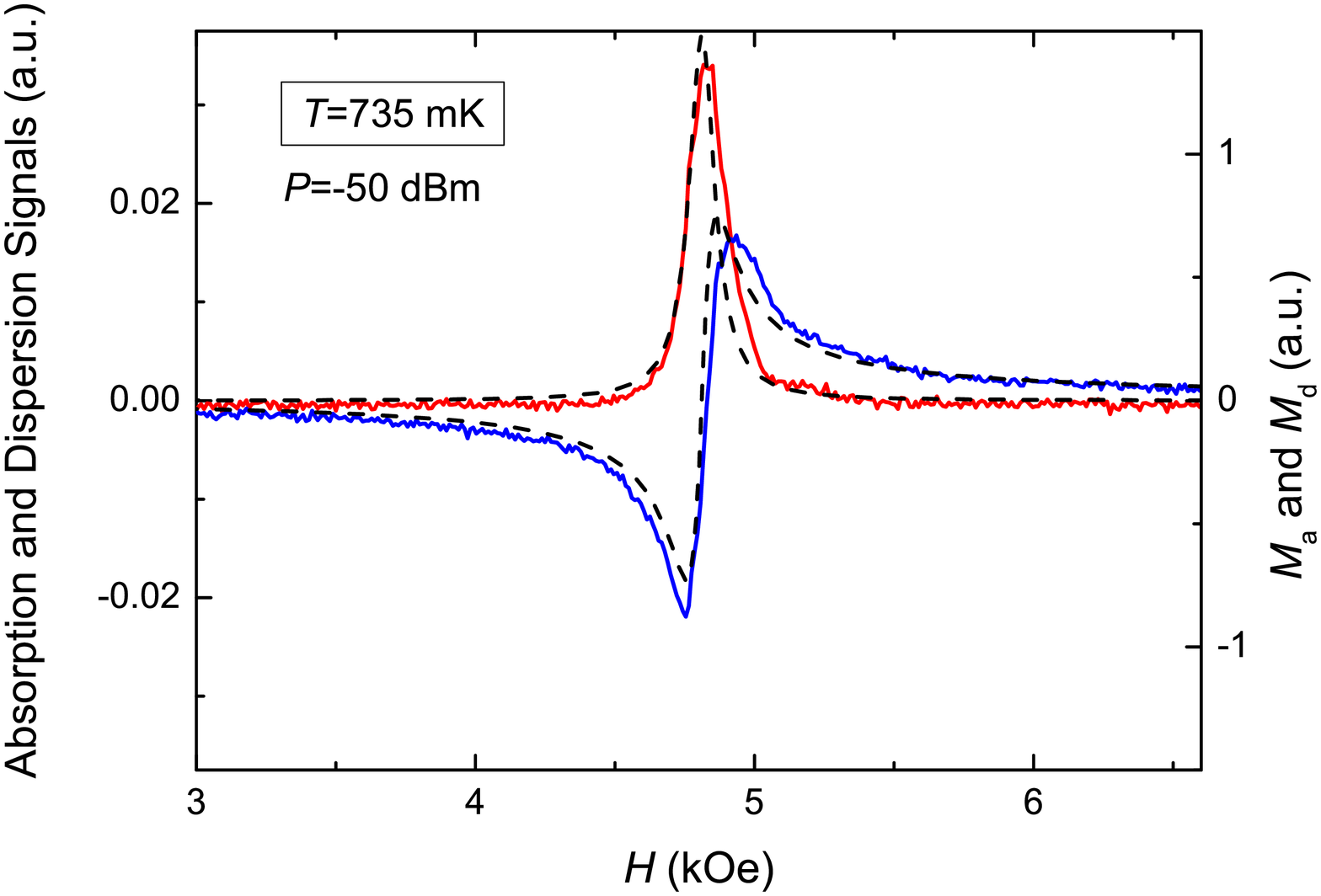}
\caption{(color online) Absorption (solid line, red) and dispersion (solid line, blue) signals measured in MnCO$_3$ at $T=735\,$mK and at low input rf power of $P=-50\,$dBm. Dashed lines are the ac magnetization components $M_a$ and $M_d$ calculated as described in Section III(C).}
\label{fig:lowP}
\end{figure}

Figure~\ref{fig:lowP} shows the NMR signals proportional to $M_a$ (absorption signal) and $M_d$ (dispersion signal) measured at $T=735\,$mK as a function of the static field $H$ at a low input rf power of $-50\,$dBm. At such low power, the deviation of nuclear magnetization from its equilibrium orientation is very small. Correspondingly, frequency pulling is very close to its value for the equilibrium spin system; therefore, nonlinearity of the NMR signal is negligible. These signals are used to find the fitting parameters that define the dependence of frequency pulling and the nuclear magnetic relaxation rate on the value of $H$. It should be noted that although these signals were measured in the linear regime of NMR, they are not ideally symmetric. This suggests that the sample used in the experiment is not a single crystal, which is also confirmed by careful visual examination under the microscope. It is likely that the sample consists of at least two or three mono-crystals of different size with somewhat different relative orientations of their lattices. 

\begin{figure}[htt]
\includegraphics[width=0.48\textwidth]{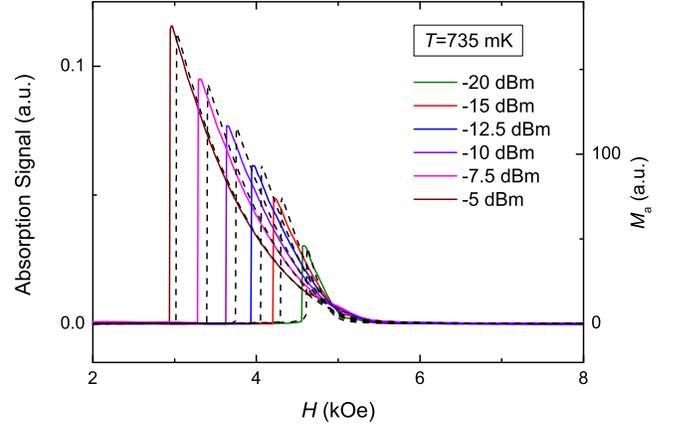}
\caption{(color online) Absorption signals (solid lines) measured in MnCO$_3$ at $T=735\,$mK and at several values of input rf power (shown in the figure) during down-field sweep. Dashed lines represent values of $M_a$ calculated as described in Section III(C).}
\label{fig:abs}
\end{figure}

\begin{figure}[htt]
\includegraphics[width=0.48\textwidth]{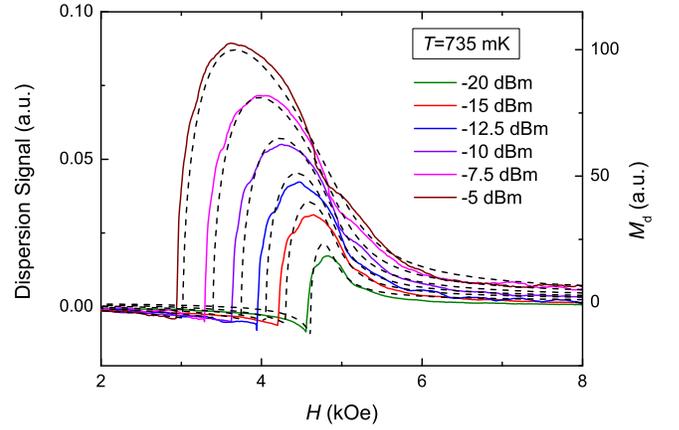}
\caption{(color online) Dispersion signals measured in MnCO$_3$ at $T=735\,$mK and at several values of input rf power (shown in the figure) during down-field sweep. Dashed lines represent values of $M_d$ calculated as described in Section III(C).}
\label{fig:disp}
\end{figure}

\begin{figure}[htt]
\includegraphics[width=0.48\textwidth]{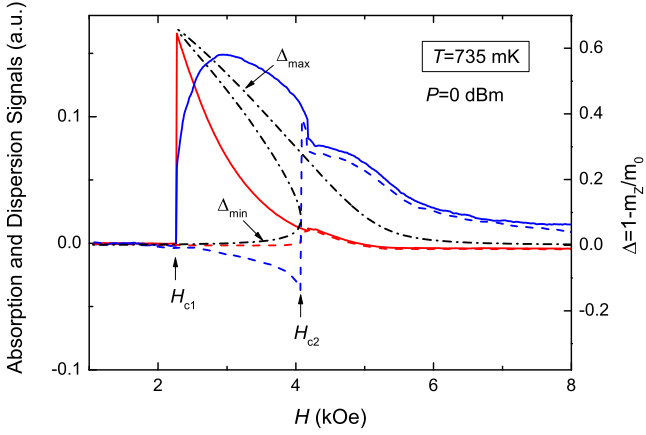}
\caption{(color online) Absorption and dispersion signals measured at $T=735\,$mK at highest input rf power of $P=0\,$dBm during up-field (dashed lines) and down-field (solid lines) sweeps. The dash-dotted line is the result of numerical calculations for $\Delta=1-m_Z/m_0$ as described in Section III(C).}
\label{fig:highpower}
\end{figure}

Figures~\ref{fig:abs} and \ref{fig:disp} show absorption and dispersion signals measured at $T=735\,$mK and at different input rf power levels. It is clear that with increasing power, the NMR frequency shifts toward lower values of $H$ due to  frequency pulling. Both absorption and dispersion signals become very asymmetrical. In addition, both signals show strong hysteresis upon reversing the direction of the field sweep. In Figure~\ref{fig:highpower} we show the absorption and dispersion signals obtained at the highest input power of $P=0\,$dBm used in this experiment. In the next section, we compare the measured absorption and dispersion signals with predictions of our theory, see Eqs.~(\ref{eq:mXstead}-\ref{eq:mZstead}). In addition, we compare the experimental results with predictions based on the phenomenological Bloch equations, see Eqs.~(\ref{eq:mXsteadBloch}-\ref{eq:mZsteadBloch}).     
%
%

\subsection{Expressions for ac magnetization components and their comparison with experimental results}

In order to compare the results presented in Figs.~(\ref{fig:abs}-\ref{fig:disp}) with predictions of our theory, first we need to find relation between the ac magnetization components $M_a$ and $M_d$ and components of nuclear magnetization $m_X$ and $m_Y$ described by Eqs.~(\ref{eq:mXstead}-\ref{eq:mYstead}). It is important to realize that due to the much larger magnitude of the electron magnetization vector $\textbf{M}=(M^{\xi},M^{\eta},M^{\zeta})$ compared to that of the nuclear magnetization vector $\textbf{m}=(m^{\xi},m^{\eta},m^{\zeta})$, the ratio of which is approximately $\gamma_e/\gamma_n \gtrsim 10^3$, the main contribution to the observed ac magnetization comes from oscillations of vector $\bf{M}$. We can find necessary expressions using the relation between the complex amplitudes of $M^{\xi}(t)$ and $m^{\xi}(t)$, as given by Eq.~(\ref{A8}-\ref{A11}) in the Appendix. 

According to Fig.~\ref{Fig:Geom} and Eqs.~(\ref{eq:sin}-\ref{eq:magMm}), the vector component of electron magnetization $\mathbf{M}_{\parallel}(t) \parallel \mathbf{h}(t) $, which defines the observed NMR signals, can be expressed in terms of $M^{\xi}(t)$ as

\begin{eqnarray}
&& M(t)=M_1^z + M_2^z = -(M_1^{x_1}+M_2^{x_2})\sin\psi = \nonumber \\
&& = -2\left( \frac{H+H_D}{H_E} \right) M^\xi (t).
\label{Eq:MObs}
\end{eqnarray}

\noindent Assuming time dependence of the rf magnetic field in the form $h(t)=2H_1\cos \omega t$, absorption $M_a$ and dispersion $M_d$ signals are given by, respectively, the imaginary and real parts of the complex amplitude of $M_{\parallel}(t)$. The expressions for $M_d$ and $M_a$ in terms of oscillating components of the nuclear magnetization vector can be found from the above equation using Eqs.~(\ref{A8}-\ref{A11}) and approximations discussed in the Appendix. Using the transformation (\ref{eq:RWT}) and expressing results in terms of nuclear magnetization vector components $m_X$ and $m_Y$ we obtain   

\begin{eqnarray}
&& M_d= M_d^{nr} +  \frac{H_n}{H} \left( 1-\frac{\omega^2}{\omega_e^2} \right)^{-1}  \left[ m_X - \frac{\omega}{\omega_n\omega_pT_{2n}}  m_Y \right], \nonumber\\
&& M_a= \frac{H_n}{H} \left( 1-\frac{\omega^2}{\omega_e^2} \right)^{-1}  \left[ m_Y + \frac{\omega}{\omega_n\omega_pT_{2n}}  m_X \right].
\label{eq:MdMa} 
\end{eqnarray}

\noindent In the above equations, the non-resonant contribution $M_d^{nr}$ to the dispersion signal is given by

\begin{equation}
M_d^{nr}=\frac{H_1 H_n}{H^2} \left( 1-\frac{\omega^2}{\omega_e^2} \right)^{-2}\frac{\omega_n}{\omega_p} m_0.
\end{equation}

\noindent It arises from the direct coupling of the electron magnetization vector to the rf magnetic field having amplitude $H_1$. The resonant contribution (at the NMR frequency) comes from the terms containing $m_X$ and $m_Y$. The main contribution to the dispersion (absorption) signal comes from the $m_X$ ($m_Y$) component of the nuclear magnetization vector. Note that it is enhanced by the amplification factor $\eta$ given by Eq.~(\ref{A13}) in the Appendix. The components $m_X$ and $m_Y$ are given by equations (\ref{eq:mXstead}-\ref{eq:mYstead}) and are expressed in terms of the dimensionless parameter $\Delta=1-m_Z/m_0$ determined from Eq.~(\ref{eq:mZstead}). It is instructive to discuss the general form of solutions of these equations. Eq.~(\ref{eq:mZstead}) is a fourth order polynomial equation with respect to $\Delta$. One of the roots of Eq.~(\ref{eq:mZstead}), viz. $\Delta\approx$2, corresponds to an unstable state with antiparallel orientation of the vectors $\bf{m}$ and $\bf{H}_{n}$. The other three roots are the solutions of a third order polynomial equation that is obtained from (\ref{eq:mZstead}) by excluding the root $\Delta\approx$2. We are interested in the real-valued roots and their dependence on $H$, which is defined by the corresponding dependencies of $\omega_{1}(H)$, $\omega_{p}(H)$ and $T_{2n}(H)$ given by Eqs.~(\ref{A12}-\ref{A15}) in the Appendix. 
%
%

At low rf excitation powers, such that $\omega_1T_{2n}\ll 1$ the equation (\ref{eq:mZstead}) has only one real root, given by

\begin{equation}
\Delta_1=\frac{\omega_1^2}{(\omega_n -\omega -\omega_p)^2+T_{2n}^{-2}} \ll 1.
\end{equation} 

\noindent Using this expression in Eqs.~(\ref{eq:mXstead}-\ref{eq:mYstead}) we obtain

\begin{eqnarray}
&&\frac{m_{X}}{m_{0}}=\frac{ ( \omega_{n}-\omega - \omega_{p} ) \omega_{1}}{( \omega_{n}-\omega-\omega_{p})^{2}+T_{2n}^{-2}}, \nonumber \\
&&\frac{m_{Y}}{m_{0}}=\frac{\omega_{1}T_{2n}^{-1}}{( \omega_{n}-\omega-\omega_{p} )^{2}+T_{2n}^{-2}}.
\label{eq:linear}
\end{eqnarray}

\noindent At fixed rf frequency $\omega$ the position of the resonance in the static magnetic field $H$ is determined by the dependence of $\omega_p$ on $H$ given by Eq.~(\ref{A14}) in the Appendix. The resonance value of field $H=H_{res}$ is defined by 

\begin{equation}
\omega_{n}-\omega-\omega_{p}(H_{res})=0.
\label{Eq:ResDef}
\end{equation}

\noindent The width of the resonance $\Delta H$, which can be defined as the distance between the maximum and minimum of the dispersion signal, is related to $T_{2n}$ as  

\begin{eqnarray}
T_{2n}^{-1}(H_{res})=-\frac{1}{2}\frac{\partial \omega_{p}(H_{res})}{\partial H}\Delta H.
\label{Eq:T2nRes}
\end{eqnarray}

\noindent The dependence of $T_{2n}$ on $H$ is given by Eq.~(\ref{A15}) in the Appendix. It is convenient to express the dependence of $\omega_p$ and $T_{2n}$ on $H$ in terms of adjustable parameters $a_1$ and $a_2$ as

\begin{eqnarray}
\label{Eq:omegap}
&& \omega_p(H)=\frac{ a_1}{H(H+H_{D})}, \\
\label{Eq:T2n}
&& T_{2n}^{-1}(H) =\frac{a_2}{H^{2}(H+H_{D})^{2}},
\end{eqnarray}

\noindent and determine the numerical values of $a_1$ and $a_2$ at a given temperature from the comparison of experimental data taken at low rf excitation power and the ac magnetization components $M_a$ and $M_d$ calculated using Eqs.~(\ref{eq:MdMa}) and (\ref{eq:linear}). Such a comparison is shown in Fig.~\ref{fig:lowP} where $M_a$ and $M_d$ calculated using ${a_1/2\pi=2.06\times 10^3\,}$MHz(kOe)$^2$ and ${a_2/2\pi=1.6\times 10^3\,}$MHz(kOe)$^4$ are plotted using dashed lines. Here, we used $H_D=4.4\,$kOe, $\omega_n/2\pi =640\,$MHz, and $\omega/2\pi =593.5\,$MHz. The value of $H_1$ was chosen to be sufficiently small to ensure a linear NMR regime. The corresponding values of $H_{res}$ and $\Delta H$ are approximately $4.81\,$kOe and $0.11\,$kOe, respectively. The width of the resonance corresponds to the nuclear relaxation $T_{2n}$ time of approximately $0.19\,\mu$s.

For arbitrarily large rf excitation powers there exists a range of $H$ where there is more than one real-valued root of Eq.~(\ref{eq:mZstead}). The numerical solution of this equation for a large value of $H_1$ is shown by the dash-dotted line in Fig.~\ref{fig:highpower}. In this calculation, we used $\omega_p(H)$ and $T_{2n}(H)$ given by Eqs.~(\ref{Eq:omegap}-\ref{Eq:T2n}) with numerical values of $a_1$ and $a_2$ given above. The interval $H_{c1}<H<H_{c2}$ corresponds to the region where three real roots exist. The merger of two real roots (where they become complex) corresponds to the phenomenon which in mathematics is called {\it catastrophe}. We restrict ourselves to the analysis of the upper branch $\Delta_{max} (H)$, which has its critical point at $H=H_{c1}$, see Fig.~\ref{fig:highpower}. The state of the lower branch $\Delta_{min}(H)$ is destroyed at fields $H<H_{c2}$ due to instabilities caused by the excitation of $n$-magnons that are in resonance with the rf pumping field.~\cite{Tulin,Kurkin2} 

In order to make comparison between the experimental data presented in Figs.~\ref{fig:abs}-\ref{fig:disp} and the ac magnetization components $M_a$ and $M_d$ calculated using Eqs.~(\ref{eq:mXstead}-\ref{eq:mZstead},\ref{eq:MdMa}), we need to establish correspondence between the values of rf excitation power used in the experiment and the numerical values of $H_1$ used in the calculations. We used the value of $H_{c_1}\approx 2.2\,$kOe in Fig.~\ref{fig:highpower} to define the value of $H_1=23\,$Oe at the excitation power of $0\,$dBm. The line shapes of $M_a$ and $M_d$ calculated for the corresponding values of $H_1$ and using Eqs.~(\ref{Eq:omegap}) and (\ref{Eq:T2n}) are plotted in Figs.~\ref{fig:abs} and \ref{fig:disp}, respectively, using dashed lines. As seen in these figures, we obtain very good quantitative agreement between our experimental data and predictions of our theoretical model based on Eqs.~(\ref{eq:mXLLG}-\ref{eq:mZLLG}). Note that conservation of the magnitude of vector $\bf{m}$, which follows from our theory, allows us to express $\Delta_{max} (H)$ in terms of the deflection angle $\beta_0$ between the vector $\bf{m}$ and its equilibrium orientation

\begin{equation}
\Delta_{max} (H) = 1- \cos \beta_0.
\end{equation}

\noindent At the critical field $H_{c1}\approx 2.2\,$kOe in Fig.~\ref{fig:highpower} corresponding to the input rf power of $0\,$dBm the deflection angel has value $\beta_0\approx 70\,$degrees.

For the sake of comparison with the predictions of the Bloch model based on equations (\ref{eq:mXBloch}-\ref{eq:mZBloch}), it is instructive to consider the dependence of the absorption signal $M_a$ at $H=H_{c1}$ on the input rf power $P$. Note that the critical field $H=H_{c_1}$ corresponds to the resonant condition 

\begin{equation}
\omega_n-\omega-\omega_p+\omega_p\Delta=0.
\label{eq:reson}
\end{equation}   

\noindent From Eqs.~(\ref{eq:mXstead}-\ref{eq:mYstead}) for the steady state solutions, this gives $m_X(H_{c_1})=0$ and $m_Y(H_{c_1})=m_0 \omega_1 T_{2n}$; therefore, from the second line of (\ref{eq:MdMa}) and Eqs.~(\ref{A12}-\ref{A15}) in the Appendix we obtain

\begin{equation}
M_a(H_{c_1})=\frac{\gamma_nm_0H_n^2}{a_2} \left( 1-\frac{\omega^2}{\omega_e^2} \right)^{-2} \left( H_{c_1} + H_D \right)^2 H_1.
\label{eq:MaHc}
\end{equation} 

\begin{figure}[htt]
\includegraphics[width=0.41\textwidth]{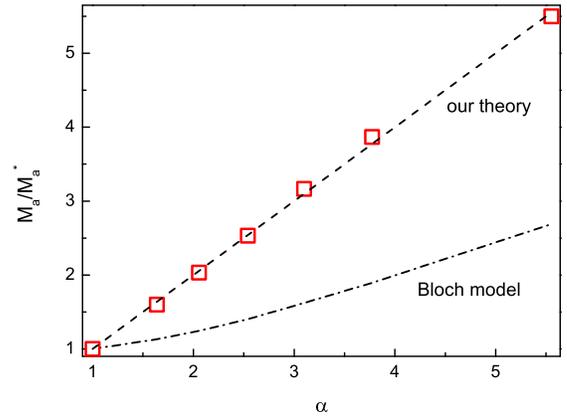}
\caption{(color online) Normalized values of the absorption signal at the critical field $H=H_{c1}$ as a function of quantity $\alpha(H_{c1},\sqrt{P/P^*})$ defined by Eq.~(\ref{eq:alpha}) obtained from the experimental data presented in Figs.~\ref{fig:abs} and \ref{fig:highpower} (opened squares). The lines are the corresponding dependences predicted by our theory (dashed line) and by the conventional Bloch approach (dash-dotted line).}
\label{fig:comp}
\end{figure}

\noindent In order to get rid of the field-independent factor in front of the right-hand side of the above equation, it is convenient to normalize this expression to the value of $M_a(H_{c1})$ at one of the values of the input rf power used in the experiment. For example, we choose the lowest power of $P^*=-20\,$dBm used in Fig.~\ref{fig:abs}. The expression for the normalized absorption signal becomes (for the sake of simplicity we neglect $\omega/\omega_e<<1$)

\begin{equation}
\frac{M_a(H_{c1})}{M_a^*(H_{c1})}\approx \left( \frac{H_{c1}+H_D}{H_{c1}^*+H_D} \right)^2 \frac{H_1}{H_1^*}=\alpha(H_{c_1},\sqrt{P/P^*}),
\label{eq:alpha}
\end{equation} 

\noindent where $M_a^*$, $H_{c1}^*$ and $H_1^*$ are the corresponding values of $M_a$, $H_{c1}$ and $H_1$ at the input rf power $P=P^*$. Note that the normalized absorption signal depends only on the values of the critical field $H_{c1}$ and the ratio $\sqrt{P/P^*}=H_1/H_1^*$, which can be readily determined in the experiment. Figure \ref{fig:comp} shows dependence of $M_a(H_{c1})/M_a^*(H_{c1})$ on $\alpha(H_{c1},\sqrt{P/P^*})$ obtained from the experimental data shown in Figs.~\ref{fig:abs} and \ref{fig:highpower} (opened squares), as well as theoretically predicted dependence given by Eq.~(\ref{eq:alpha}) (dashed line). 

Now we compare this result with predictions of the heating scenario based on the phenomenological Bloch equations (\ref{eq:mXBloch}-\ref{eq:mZBloch}). Note that as in Eq.~(\ref{eq:mZstead}), the third order polynomial equation (\ref{eq:mZsteadBloch}) with respect to $\Delta$ can have three real-valued roots in certain ranges of the static magnetic field $H_{c1}<H<H_{c2}$. This may result in nonlinear NMR signals similar to those described earlier.~\cite{deGennes} Using condition (\ref{eq:reson}) satisfied at the field $H=H_{c1}$, we obtain from Eqs.~(\ref{eq:mXsteadBloch}-\ref{eq:mZsteadBloch})

\begin{eqnarray}
&& m_x(H_{c1})=0, \quad m_y(H_{c1})=\omega_1T_2 m_z(H_{c1}), \nonumber \\
&& \quad m_Z(H_{c1})=\frac{m_0}{1+\omega_1^2T_1T_2}.
\label{eq:BlochHc}
\end{eqnarray} 

\noindent Note that, unlike in our proposed theory, the Bloch model does not conserve the magnitude of the magnetization vector. Indeed, from Eq. (\ref{eq:BlochHc}) we have 

\begin{equation}
|\textbf{m}| = \sqrt{\frac{1+\omega_1^2T_2^2}{\left( 1 + \omega_1^2T_1T_2 \right)^2 } } m_0.
\label{eq:absm}
\end{equation}

\noindent Typically, we have $T_2\ll T_1$; therefore, $|\textbf{m}|$ can already be significantly less than $m_0$ for $\omega_1^2 T_1 T_2 \gtrsim 1$. 

Using Eqs.~(\ref{eq:reson},\ref{eq:BlochHc}) it is convenient to write $\omega_1^2T_1T_2$ as 

\begin{equation}
\omega_1^2T_1T_2 = \frac{\Delta\omega(H_{c1})}{\omega_p(H_{c1}) - \Delta\omega(H_{c1})},
\label{eq:omega1T1T2}
\end{equation}

\noindent where $\Delta\omega=\omega - \omega_n +\omega_p$. Note that using the above equation, the values of $\omega_1^2 T_1 T_2$ at $H=H_{c1}$ can be determined from the experimental values of $H_{c_1}$. Using the second line of Eq.~(\ref{eq:MdMa}) we obtain the expression for the absorption signal at the catastrophe field

\begin{equation}
M_a(H_{c_1})=\frac{m_0H_n}{H}\sqrt{\frac{T_2}{T_1}} \left( 1-\frac{\omega^2}{\omega_e^2} \right)^{-1} \frac{\sqrt{\omega_1^2T_1T_2}}{1+\omega_1^2T_1T_2},
\label{eq:MaHcBloch}
\end{equation} 

\noindent and using Eqs. (\ref{A12}-\ref{A13}) from the Appendix and the definition of $\alpha$, see Eq.~(\ref{eq:alpha}), we obtain for the normalized absorption signal

\begin{equation}
\frac{M_a(H_{c1})}{M_a^*(H_{c1})}\approx \left( \frac{H_{c1}^*(H_{c1}^*+H_D)}{H_{c1}(H_{c1}+H_D)} \right)^2 \frac{1+(\omega_1^2 T_1T_2)^*}{1+\omega_1^2 T_1T_2} \alpha,
\label{eq:alphaBloch}
\end{equation} 

\noindent where $\left( \omega_1^2 T_1 T_2\right)^*$ is the value of $\omega_1^2 T_1 T_2$ at the critical field for the input rf power $P=P^*$. Again, note that the term in front of $\alpha$ on the right side of the above equation depends only on the value of the critical field $H=H_{c1}$, which can be determined from the experiment, and does not depend on the choice of values for $T_1$ and $T_2$. The dependence of the normalized absorption signal on $\alpha$ given by Eq.~(\ref{eq:alphaBloch}) is plotted in Fig.~\ref{fig:comp} with a dash-dotted line. Clearly, the conventional Bloch approach fails to adequately account for the behavior observed in the experiment. Using Eq.~(\ref{eq:omega1T1T2}), it is straightforward to estimate the value of $\omega_1^2T_1T_2$ at the critical field $H=H_{c1}$ for the maximum input rf power $P=0\,$dBm. Using the value of $H_{c1}\approx 2.2\,$kOe, see Fig.~\ref{fig:highpower}, we obtain $\omega_1^2 T_1 T_2\approx 2$. Thus, the Bloch theory predicts a significant reduction of the magnitude of the nuclear magnetization vector due to heating. This does not agree with our experimental results.  

\section{Discussion}

The mechanism of indirect relaxation of nuclear spins via the electron subsystem, which preserves the magnitude of nuclear magnetization, has important consequences for the dynamics of n-magnons under external rf pumping. In this section we provide some detailed discussion of the processes involved and show that the pumping of n-magnons with a uniform ac magnetic field $\textbf{h}(t)$ in a wide range of frequencies results in a macroscopic accumulation of n-magnons with $\textbf{k}=0$, which can be identified with the BEC of nonequilibrium n-magnons.

\begin{figure}[htt]
\includegraphics[width=0.45\textwidth]{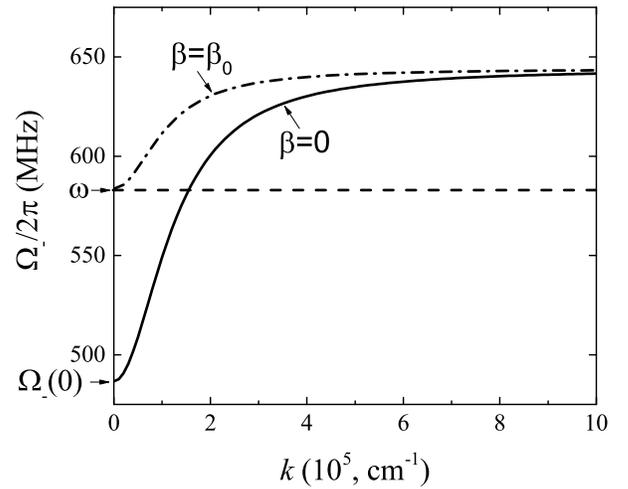}
\caption{The spectrum of n-magnons in $\mathrm{MnCO_{3}}$ for zero deflection angle (solid line) and for deflection angle given by Eq.~(\ref{eq:beta0}) (dash-dotted line). The frequency of rf pumping $\omega$ is indicated by a dashed line.}
\label{fig:magnspectr}
\end{figure}

As pointed out earlier,~\cite{Kurkin2} an external rf pumping of the coupled electron-nuclear spin system with a uniform $\textbf{h}$-field at the frequency $\omega$ in the range $\Omega_-(0)< \omega < \omega_n$ can lead to instability against formation of n-magnons with $k\neq 0$ corresponding to $\Omega_-(k)=\omega$, see Fig.~\ref{fig:magnspectr}. Appearance of such magnons slightly decreases the deflection angle $\beta$ of the nuclear magnetization vector from its equilibrium orientation. Further time evolution of such magnons depends on the nuclear relaxation processes. According to the Bloch approach scenario, such magnons thermalize and lead to an temperature increase of the nuclear system $T$. This would significantly decrease the magnitude of nuclear magnetization $m_0(T)$ and modify the spectrum of n-magnons such that the frequency of the $k=0$ mode, given by Eq.~(\ref{eq:bot}), coincides with $\omega$. A stationary state of the nuclear spin system would correspond to a small deflection angle $\beta=\arccos (m_z/m_0(T))$.  As we showed above, such conventional heating scenario does not agree with the experiments described here. 

In an alternative scenario, two $n$-magnons with wave vectors $\pm \textbf{k}$ can mutually transform their $\left |\bf{k} \right |$ in momentum-conserving, four-magnon processes to the value $\left | \bf{k} - \delta\bf{k} \right |$, which corresponds to the new resonance condition $\Omega_{-} (\left | \bf{k} - \delta \bf{k} \right |,\beta)=\omega$. Thus, the rf pumping and accumulation of magnons continue, leading to a further increase of the deflection angle $\beta$, and so on. The accumulation of magnons and increase of the deflection angle stops when the bottom of the frequency band of n-magnons reaches the rf pumping frequency

\begin{equation}
\omega_{n}-\omega_{p}\cos\beta_0=\omega,
\label{eq:beta0}
\end{equation}

\noindent see dash-dotted line in Fig.~\ref{fig:magnspectr}. In this scenario, all accumulated n-magnons, which provide the value of the deflection angle in the above equation, are transformed into n-magnons with $k=0$. Such a macroscopic accumulation of magnons in a single $k=0$ mode can be identified as a Bose-Einstein condensate of nuclear spin waves by analogy with the atomic BEC. In this state of the nuclear system, the vector of nuclear magnetization is deflected by a large angle $\beta_0$ from the equilibrium orientation (e.g. as large as 70~degrees in the experiments described here), while the magnitude of the magnetization vector is conserved. This description agrees with the interpretation of experimental results observed earlier.~\cite{Kazan1,Kazan2,Kazan3}

Investigation of the coupled electron-nuclear spin precession in easy-plane antiferromagnets is interesting in the context of conventional magnon BEC. Unlike the conventional atomic BEC obtained by cooling the atomic system, the magnon BEC is established by an external rf pumping that compensates for the loss of quasiparticles. The magnon BEC was first observed in the antiferromagnetic B-phase of superfluid $^3$He.~\cite{experiment,theory1} Owing to its absolute purity and the specific form of the magnetic energy potential, which ensures stability of BEC, this system became a test-bed for investigation of conventional magnon BEC with $k=0$ showing close analogy to the atomic BEC.~\cite{book,BECanalog} Later, the stable state of magnon BEC was also obtained in antiferromagnetic A-phase of superfluid $^3$He immersed in aerogel.~\cite{BECJ,BECG} Superfluid $^3$He-A is particularly interesting in the context of the work presented here. This system is a two-sublattice antiferromagnetic quantum liquid in which the magnetic part of the Hamiltonian exactly corresponds to that of the easy-plane solid antiferromagnets considered here.~\cite{Leggett} For this reason, the latter systems were suggested for observation of conventional magnon BEC.~\cite{BunkovUFN} The significant advantage of these systems over superfluid $^3$He is that they require rather moderate cryogenic temperatures around 1~K for observation and study of BEC properties and spin superfluidity. We should point out that the conventional magnon BEC with $k=0$ discussed here has very different properties from those of e-magnon BEC observed in ferromagnetic yttrium iron garnet (YIG) films.~\cite{YIGBEC,Hillebrant} In the latter case, the minimum of the magnetic energy corresponds to spin waves with $k\neq 0$ and has some similarities with charged density waves in superconductors.      

\section{Conclusions}

We present a description of coupled electron-nuclear spin systems using the coupled Landau-Lifshitz-Gilbert equations, which take into account an indirect relaxation of nuclear spins via the electron spin subsystem. In our theory, the magnitude of the nuclear magnetization is conserved for arbitrary large excitation powers, which is drastically different from the conventional heating scenario based on the phenomenological Bloch approach. In particular, we consider the coupled electron-nuclear spin motion in the easy-plane antiferromagnetic crystals with a weak magnetic anisotropy in the basal plain. The equations of motion are solved analytically, assuming small deviations of electron magnetization vectors from their equilibrium orientations, which allow us to linearize the corresponding equations of motion for the electron subsystem.  The solutions obtained are compared with experimental nonlinear NMR signals obtained in a MnCO$_3$ sample at temperatures below 1~K,  and good quantitative agreement is found. This suggests  that at high excitation rf powers used in the experiment, the nuclear magnetization vector can be deflected at very large angles, exceeding 70 degrees. This result is drastically different from predictions of the standard heating scenario based on the phenomenological Bloch approach. In particular, the latter predicts significant reduction of the magnitude of the nuclear magnetization vector under conditions of our experiment. The proposed theory brings together properties of nonlinear NMR in magnetic systems in solids considered here and the superfluid, ${^3}$where the magnitude of the nuclear magnetization vector is also conserved. This provides the theoretical background for the study of BEC of nonequilibrium magnons in $\mathrm{MnCO_{3}}$.

{\bf Acknowledgements}
The work of L.~V.~A and D.~K. was supported by an internal grant from the Okinawa Institute of Science and Technology Graduate University and JSPS KAKENHI (Grant No. 26400340). Part of the work is done by the order of FASO of the RF: subject "Spin" No 01201463332, grant No 15-9-2-49; subject "Electron No 01201463326, grant No 15-8-2-10. Yu.~M.~B acknowledges support from the Okinawa Institute of Science and Technology Graduate University as a visiting scientist. We thank Jason Ball and Steven Aird for proof-reading the manuscript.

\appendix*

\section{Linearized equations for electron spin precession}

The system of coupled equations (\ref{eq:M}-\ref{eq:m}) contains twelve dynamical variables corresponding to three components for each of four magnetization vectors $\mathbf{M}_{1,2}$ and $\mathbf{m}_{1,2}$. This system conveniently splits into two independent subsystems, each consisting of six equations, if we introduce new dynamical variables according to~\cite{theory}

\begin{equation}
2M^{\alpha_{\pm}}= M_{1}^{\alpha_{1}} \pm M_{2}^{\alpha_{2}}, \quad 2m^{\alpha_{\pm}}= m_{1}^{\alpha_{1}} \pm m_{2}^{\alpha_{2}},
\label{A1}
\end{equation}

\noindent where the upper indexes are defined as $\alpha_{\pm} \in \left [x_{\pm},y_{\pm},z_{\pm} \right ]$, $\alpha_{1} \in \left [x_{1}, y_{1}, z_{1} \right ]$, $\alpha_{2} \in \left [x_{2}, y_{2}, z_{2} \right ]$, and $M_i^{x_i}$, $M_i^{y_i}$, $M_i^{z_i}$ ($m_i^{x_i}$, $m_i^{y_i}$, $m_i^{z_i}$), $i=1,2$, are components of electron (nuclear) magnetization vectors in the corresponding reference frames $(x_1,y_1,z_1)$ and $(x_2,y_2,z_2)$ defined in Fig.~\ref{Fig:Geom}. The components $M^{\alpha_{-}}$ correspond to the high-frequency branch of oscillations of the vectors $\textbf{M}_{1,2}$, which is weakly affected by the interaction between electron and nuclear spins. For this reason, only the low-frequency branch of $\textbf{M}_{1,2}$ is of interest for NMR. This branch is described by the components $M^{\alpha_{+}}$ and $m^{\alpha_{+}}$. It is convenient to mark the indexes $x_+$, $y_+$ and $z_+$, respectively, as $\xi$, $\eta$ and $\zeta$. In what follows, we consider the static magnetic fields $H>H_c$. The field $H_c$ is defined by the relation
\begin{eqnarray}\label{A2}
H_c(H_c+H_D) = H_EH_nm_0/M_0,\ H_n = AM_0,
\end{eqnarray}
where $M_0$ and $m_0$ are the magnitudes of vectors $\mathbf{M_1},\mathbf{M_2}$ and $\mathbf{m_1},\mathbf{m_2}$ respectively. The inequality ${H>H_c}$ is responsible for conditions
\begin{equation}
M^{\xi,\eta}\ll M_0, \quad M^{\zeta}\approx M_0,
\label{A3}
\end{equation}
In this approximation, linearized equations of motion for $M^\xi$ and $M^\eta$ become~\cite{theory} 
\begin{eqnarray}
&& \frac{1}{\gamma_e}\frac{\mathrm{d}M^{\xi}}{\mathrm{d}t}=\left( H_E+H_A \right) M^{\eta}-H_n m^{\eta}-\frac{1}{\gamma_e T_{2e}}M^{\xi}, \nonumber \\
&& \frac{1}{\gamma_e }\frac{\mathrm{d}M^{\eta}}{\mathrm{d}t}=-\frac{H\left( H+H_D \right)}{H_E} M^{\xi} \nonumber \\
&& + \frac{H+H_D}{H_E} M_0 h(t)+H_{n}m^{\xi} - \frac{1}{\gamma_e T_{2e}} M^{\eta},
\label{A4}
\end{eqnarray}

\noindent where $T_{2e}$ is the transverse relaxation time of the electron magnetization. We will use the applied rf magnetic field $\textbf{h} \parallel \bf{z}$ in the form
\begin{equation}
h(t)=2H_{1}\cos \omega t.
\label{A5}
\end{equation}
Note that the approximation~(\ref{A3}) used to obtain the linearized Eqs.~(\ref{A4}) corresponds to the following condition
%
\begin{equation}
\left | \omega - \omega_e \right | T_{2e}\gg 1, 
\label{A6}
\end{equation}
%
where $\omega_e={\gamma_e (H(H+H_D))^{1/2}}$ is the frequency of uniform precession of vector $\textbf{M}$ neglecting the interaction between electron and nuclear spins. The condition~(\ref{A6}) is well satisfied for typical rf frequencies close to the NMR frequency.

Eq.~\eqref{A4} can be simplified further. Noting that, since $M^{\eta}\gg m^{\eta}$ and $\gamma_e(H_E+H_A)T_{2e}\gg 1$, we can neglect the last two terms in the right-hand side of the first equation in~(\ref{A4}). Moreover, since typically $H_E\sim10^5$-$10^7\,$Oe and ${H_A\sim10^2}$-$10^4\,$Oe, we can neglect $H_A$ comparing with $H_E$. Using these approximations, we obtain from (\ref{A4})

\begin{eqnarray}\nonumber
&&\frac{d^2M^{\xi}}{dt^2} + \frac{2}{T_{2e}}\frac{dM^{\xi}}{dt} + \omega_e^2M^{\xi} = 2\gamma_e^2(H+H_D)M_0H_1\cos\omega t \\\label{A7}
&&\qquad+ \gamma_e^2H_nH_Em^{\xi}
\end{eqnarray}

%
%

\noindent The above equation is solved by using the Fourier transformations

\begin{eqnarray}
M_\Omega^{\xi} = \int dt\, M^{\xi}(t)e^{i\Omega t}, \quad  m_\Omega^{\xi} = \int dt\, m^{\xi}(t)e^{i\Omega t},
\label{A8}
\end{eqnarray} 

\noindent for which we obtain

\begin{equation}
M_{\Omega}^{\xi}\approx M_{\Omega}^{\xi}(H_{1})+M_{\Omega}^{\xi}(m_{\Omega}^{\xi}),
\label{A9}
\end{equation}

\begin{equation}
M_{\Omega}^{\xi}(H_{1})=\frac{M_{0}}{\omega_{n}}\omega_{1}(\delta(\Omega+\omega) + \delta(\Omega-\omega) ),
\label{A10}
\end{equation}

\begin{equation}
M_{\Omega}^{\xi}(m_{\Omega}^{\xi})=\frac{2M_{0}}{\omega_{n}}\left( \omega_{p} - i\frac{\Omega}{T_{2n}\omega_{n}}\right) \frac{m_{\Omega}^{\xi}}{m_{0}},
\label{A11}
\end{equation}

\noindent where for the sake of convenience we use the following definitions

\begin{equation}
\omega_{1}=\gamma_{n}\eta H_{1},
\label{A12}
\end{equation}

\begin{equation}
\eta=\frac{H_n}{H}\left(1-\frac{\Omega^{2}}{\omega_{e}^{2}}\right)^{-1},
\label{A13}
\end{equation}

\begin{equation}
\omega_{p}=\frac{\omega_n}{2}\frac{H_{n}H_{E}}{H\left(H+H_{D}\right)}\frac{m_{0}}{M_{0}}\left(1-\frac{\Omega^{2}}{\omega_{e}^{2}}\right)^{-1},
\label{A14}
\end{equation}

\begin{equation}
\frac{1}{T_{2n}}=\frac{2\omega_{n}\omega_{p}}{T_{2e}\omega_{e}^{2}}\left(1-\frac{\Omega^{2}}{\omega_{e}^{2}}\right)^{-1}.
\label{A15}
\end{equation}

\noindent To obtain the above equations, we used expansion in small parameter $\Omega/T_{2e}\ll (\omega_e^2 - \Omega^2)$. The influence of the relaxation is described by the imaginary part of $M_{\Omega}^{\xi}$ in (\ref{A11}). Note that the above equations satisfy the condition $\textrm{Im} ( M_{\Omega=0}^{\xi} )=0$ which means the absence of dissipation at $\Omega=0$. It is important to fulfill this strict requirement imposed by the theory of irreversible processes.

%



The equations for the time-dependent components $m^{\xi}(t)$, $m^{\eta}(t)$ and $m^{\zeta}(t)$ of the nuclear magnetization vector can be obtained from Eqs.~(\ref{eq:m}) if we exclude the corresponding time-dependent components of the electron magnetization vector. This can be done using the obtained relation between the Fourier components $M_{\Omega}^{\xi}$ and $m_{\Omega}^{\xi}$ given by (\ref{A8}-\ref{A11}). The resulting integro-differential equations are rather complicated and can be significantly simplified using the following approximations. First, we neglect the relaxation terms in Eqs.~\eqref{eq:m}. This approximation is justified by extremely large value of the hyperfine field $\textbf{H}_{m_{1,2}}$ in the systems considered here. This approximation, in the spirit of the Landau-Lifshitz-Gilbert approach, is responsible for conservation of the magnitude of the nuclear magnetization vector $m_{0}^2=m^{\xi}(t)^2+m^{\eta}(t)^2+m^{\zeta}(t)^2$. Second, we neglect the dependence of $\eta$, $\omega_p$ and $T_{2n}$ on $\Omega$ given by (\ref{A13}-\ref{A15}). This is justified for typical frequencies of nuclear spin precession, which are much smaller than the frequency of electron spin precession $\omega_e$. Finally, we neglect harmonics terms at frequencies $2\omega,\,3\omega,\ldots$ in equations for the time-dependent components $m_X(t)$, $m_Y(t)$ and~$m_Z(t)$ defined by Eqs.~(\ref{eq:RWT}). These approximations result in Eqs.~(\ref{eq:mXLLG}-\ref{eq:mZLLG}), solutions of which are discussed in the text.


\begin{thebibliography}{99}

\bibitem{book} Yu.~M. Bunkov and G.~E. Volovik, in {\it Novel Superfluids}, eds. K.~H. Bennemann and
J.~B. Ketterson (Oxford University Press, 2013), Chap. 4.

\bibitem{Safonov} V.~L. Safonov, {\it Nonequilibrium magnons: theory, experiments and applications} (Wiley-VCH, Verlag, 2013).

\bibitem{Vasiliev2015} O. Vainio, J. Ahokas, J. Järvinen, L. Lehtonen, S. Novotny, S. Sheludiakov, K.-A. Suominen, S. Vasiliev, D. Zvezdov, V.~V. Khmelenko, and D.~M. Lee, Phys. Rev. Lett. \textbf{114}, 125304 (2015).

\bibitem{deGennes} P.~G. de Gennes, P.~A. Pinkus, F. Hartmann-Boutron, and J.~M. Winter, Phys. Rev. {\bf129}, 1105 (1963).

\bibitem{SN} H. Suhl, Phys. Rev. {\bf 109}, 606 (1958); T. Nakamura, Prog. Theor. Phys. (Kyoto) {\bf 20}, 542, (1958).

\bibitem{TurovKuleev} E.~A. Turov, V.~G. Kuleev, Sov. Phys. JETP {\bf 22}, 176 (1966). 

\bibitem{Kazan1} Yu.~M. Bunkov, E.~M. Alakshin, R.~R. Gazizulin, A.~V. Klochkov, V.~V. Kuzmin, T.~R. Safin, M.~S. Tagirov, JETP Letters  {\bf 94}, 68 (2011).

\bibitem{Kazan2} Yu.~M. Bunkov, E.~M. Alakshin, R.~R. Gazizulin, A.~V. Klochkov, V.~V. Kuzmin, V.~S. L'vov, M.~S. Tagirov, Phys. Rev. Lett. {\bf 108}, 177002 (2012).

\bibitem{Kazan3} E.~M. Alakshin, Yu. M. Bunkov, R.~R. Gazizulin et al., Journal of Physics: Conference Series {\bf 568}, 042001 (2014).

\bibitem{landau1} A.~I. Akhiezer, V.~G. Bar'yakhtar, S.~V. Peletminskii, {\it Spin Waves} (North-Holland, Amsterdam, New York, 1968).

\bibitem{landau2} E.~A. Turov, A.~V. Kolchanov, M.~I. Kurkin, I.~F. Mirsaev, V.~V. Nikolaev, {\it Symmetry and physical properties of antiferromagnetics} (CISP - Cambridge Int. Science Publishing, 2010).

\bibitem{landau3} S.~V. Vonosovskii, {\it Magnetism} (Wiley, New York, 1974).

\bibitem{tyablikov} S.~V. Tyablikov, {\it Methods in the Quantum Theory of Magnetism} (Springer, 2013).

\bibitem{theory} E.~A. Turov, M.~I. Kurkin, V.~V. Nikolaev, Sov. Phys. JETP {\bf 37}, 147 (1973).

\bibitem{Turov} E.~A. Turov, M.~P. Petrov, {\it Nuclear Magnetic Resonance in Ferro- and Antiferromagnets} (Halsted Press, New York, 1972).

\bibitem{Abragam} A. Abragam, {\it The Principles of Nuclear Magnetism} (Oxford University Press, New Your, 1961).

\bibitem{HardySR} W.~N. Hardy and L.~A. Whitehead, Rev. Sci. Instrum. \textbf{52}, 213 (1981).

\bibitem{Abdur2015} L.~V. Abdurakhimov, D. Konstantinov, Yu.~M. Bunkov, Phys. Rev. Lett. {\bf 114}, 226402 (2015).

\bibitem{Tulin} V.~A. Tulin, Sov. Phys. JETP {\bf 78}, 149 (1980).

\bibitem{Kurkin2} M.~I. Kurkin, Yu.~G. Raidugin, V.~N. Sedyshkin, A.~P. Tankeev, Sov. Phys. Sol. State {\bf 32}, 923 (1990) [Fizika Tverdogo Tela 32, 1577 (1990)].

\bibitem{experiment} A.~S. Borovik-Romanov, Yu.~M. Bunkov, V.~V. Dmitriev, Yu.~M. Mukharskii, JETP Lett. {\bf40}, 1033 (1984).

\bibitem{theory1} I.~A. Fomin, JETP Lett. {\bf40}, 1037 (1984).

\bibitem{BECanalog} Yu.~M. Bunkov, J. Low Temp. Phys., {\bf  183}, DOI 10.1007/s 10909-016-1583-z (2016).

\bibitem{BECJ} T. Sato, T. Kunimatsu, K. Izumina, A. Matsubara, M. Kubota, T. Mizusaki, Yu.~M. Bunkov, Phys. Rev. Lett. {\bf 101}, 055301 (2008).

\bibitem{BECG} P. Hunger, Yu.M. Bunkov, E. Collin, H. Godfrin, J. Low Temp. Phys. {\bf 158}, 129 (2010).

\bibitem{Leggett} A.~J. Leggett, Rev. Mod. Phys. {\bf 47}, 331 (1975).

\bibitem{BunkovUFN} Yu.~M. Bunkov, Physics-Uspekhi {\bf 53}, 843 (2010).

\bibitem{YIGBEC} S.~O. Demokritov, V.~E. Demidov, O.~Dzyapko, G.~A. Melkov, A.~A. Serga, B. Hillebrands, and A.~N. Slavin, Nature \textbf{443}, 430 (2006).

\bibitem{Hillebrant} D.~A. Bozhko, A.~A. Serga, P. Clausen, V.~I. Vasyuchka, F. Heussner,
G.~A. Melkov, A. Pomyalov, V.~S. L'vov, B. Hillebrands, Nature Physics {\bf 12}, 1057 (2016).

\end{thebibliography}
\end{document}